\documentclass[prd,preprintnumbers,nofootinbib,floatfix]{revtex4}

\usepackage{amsmath}

\begin{document}


\title{Next-to-leading order $pp\to W^\prime\to tb$ production at 14 TeV and 33 TeV}

\author{Zack~Sullivan}
\email{Zack.Sullivan@IIT.edu}
\affiliation{Department of Physics, Illinois Institute of Technology, Chicago, Illinois 60616-3793, USA}

\preprint{IIT-CAPP-13-07}

\begin{abstract}
  I update the predicted leading order and next-to-leading order cross
  sections and total widths for $W^\prime$ bosons that decay to top
  and bottom quarks at 14~TeV and 33~TeV $pp$ colliders ($pp\to
  W^\prime \to tb$).  Separate tables are included for right- and
  left-handed bosons.  Theoretical uncertainties are completely
  dominated by parton distribution function uncertainties, and are
  computed for $W^\prime_\pm$ production at a 14~TeV $pp$ collider.
\end{abstract}

\date{August 15, 2013}

\maketitle

New charged vector currents, generically called $W^\prime$ bosons,
appear in many models with extended gauge explanations of electroweak
symmetry breaking.  As long as these new bosons couple to fermions, a
model-independent search can probe the parameter space of all such
models by looking for the decay of the $W^\prime$ boson into a final
state involving a top quark and bottom quark \cite{Sullivan:2002jt}.
While the general case of mixed right- and left-handed couplings can
be probed, nearly all models assume a term enters the Lagrangian of
the form
\begin{equation}
  \mathcal{L}=\frac{g^\prime}{2\sqrt{2}} V^\prime_{ij} W^\prime_\mu 
  \bar f^i \gamma^\mu (1\pm\gamma_5) f^j+\mathrm{H.c.}\,,
\end{equation}
where $g^\prime$ plays the role of $g_{\mathrm{SM}}$ from the
Standard Model (SM), and $V^\prime_{ij}$ is a generalized
Cabibbo-Kobayashi-Maskawa (CKM) matrix that also allows for lepton
generation mixing.

Searches for $W^\prime$ bosons decaying to a top and bottom quark
final state \cite{Sullivan:2002jt} have been performed since Run~I of
the Fermilab Tevatron \cite{Acosta:2002nu}.  The now standard analysis
involves searching for a $l\nu bj$ invariant mass peak, where each
object is isolated, and applies loose $W$ and top-mass constraints to
reduce backgrounds \cite{Sullivan:2002jt}.  Last year we updated the
standard analysis to optimize the search for the 7~TeV and 8~TeV runs
of the Large Hadron Collider \cite{Duffty:2012rf}.  The ATLAS
\cite{Aad:2012ej} and CMS \cite{:2012sc} Collaborations used the cross
sections of Ref.\ \cite{Duffty:2012rf} to publish strong lower bounds
on the mass of a $W^\prime$ boson.  Current mass limits are $\sim
2000$~GeV if the $W^\prime$ boson has Standard Model-like (SM)
couplings.

A previous full simulation study of the reach at a 14~TeV $pp$
collider was performed in 2003 \cite{Sullivan:2003xy}.  The
conclusions of that study showed the mass reach with SM-like couplings
should be $\sim 5$--$5.5$~TeV, or about $1/3$ of the collider energy.
This limit is directly attributable to the loss of quark-antiquark
luminosity in a $pp$ collider at large-$x$ proton momentum fraction.
More important, Ref.\ \cite{Sullivan:2003xy} presented the
model-independent reach in the coupling ratio
$g^\prime/g_{\mathrm{SM}}$ versus $W^\prime$ mass, and compared it to
a few classes of models.  One of the key results of Ref.\
\cite{Sullivan:2002jt} was that all models with narrow-width
$W^\prime$ bosons fall somewhere in this space.  Hence, all
experimental measurements should produce limits, or a discovery, as a
function of $g^\prime/g_{\mathrm{SM}}$ versus $M_{W^\prime}$.

Just prior to the final meeting of the Community Summer Study, in
Ref.\ \cite{Duffty:2013aba} we proposed a new analysis that expands
the reach in the $tb$ final state by searching for a ``boosted-top''
tag, where the $Wj$ decay products of the top quark decay can not be
isolated.  In addition, we introduced a new ``boosted-bottom'' tag to
identify bottom jets in the TeV range in order to overcome dijet
backgrounds.  This new search is optimized for TeV dijets with one
boosted-top and one boosted-bottom tag.  Using this search method it
should be possible to extend the existing mass reach for $W^\prime$
bosons with SM-like couplings to about $2600$ GeV, and the
model-independent reach in coupling ratio $g^\prime/g_{\mathrm{SM}}$
by nearly a factor of two if $M_{W^\prime} >1500$~GeV with existing
8~TeV LHC data.  This analysis should be done for the upcoming 14~TeV
run of the LHC as well.

This white paper summarizes the leading order (LO) and next-to-leading
order (NLO) cross sections and widths for $pp\to W^\prime\to tb$
production for the upcoming 14~TeV run of the Large Hadron Collider,
and the proposed 33~TeV run examined by the Community Summer Study
(CSS).  Detailed study of the kinematics has shown that in both the
standard \cite{Duffty:2012rf} and boosted \cite{Duffty:2013aba}
analyses, the NLO correction can be treated as a $K$-factor times LO.
Hence, the tables below list both LO and NLO results for a wide range
of possible $W^\prime$ masses.

All cross sections are calculated using the code from Ref.\
\cite{Sullivan:2002jt}, updated for modern parton distribution
function (PDF) sets and CKM matrix values.  Leading order cross
sections are calculated with CTEQ6L1 parton distribution functions
(PDFs) \cite{Pumplin:2002vw}, while NLO cross sections use CT10
\cite{Lai:2010vv} PDFs.  The top-quark mass is taken to be $173.2\pm
1$~GeV \cite{Aaltonen:2012ra}, and $\alpha_s=0.118\pm 0.001$.
Uncertainties due to the top-quark mass and $\alpha_s$ are negligible
($<1\%$) for the $W^\prime$ masses shown.  Estimates of the NLO
theoretical uncertainty due to choice of scale is also comparatively
small (1--2\%) in all cases.  Hence, the central values of the
predictions are shown to three decimal places.  Using the standard
CTEQ Modified Tolerance Method for error determination (first
published for this cross section \cite{Sullivan:2002jt}) and CT10,
the current PDF uncertainty ranges from 10--30\% depending on the
$W^\prime$ mass, and whether it is right- or left-handed.  These
uncertainties should come down with improved fitting of large-$x$ PDFs
using LHC data.

Cross sections are calculated for $t\bar b$ and $\bar t b$ separately,
for both 14~TeV and 33~TeV $pp$ colliders.  Cross sections at 14~TeV
for right-handed $W^\prime_R$ bosons are listed by mass at LO and NLO
in femtobarns in Table \ref{tab:sigwprf}.  Current PDF uncertainties
are shown for the NLO cross sections.  Cross sections for left-handed
$W^\prime_L$ bosons are listed by mass at LO and NLO in femtobarns in
Table \ref{tab:sigwplf}.  Note, the left-handed cross sections assume
\textit{no interference} with the standard model production process.
See Sec.\ II of Ref.\ \cite{Duffty:2012rf} for a description of how to
use the left-handed normalizations.  Cross sections at 33~TeV are
listed for $W^\prime_R$ and $W^\prime_L$ in Tables \ref{tab:sigwprt}
and \ref{tab:sigwplt}, respectively.  LO and NLO total widths for the
$W^\prime$ bosons are also provided in all tables for use in
simulation studies.

\begingroup
\renewcommand{\arraystretch}{2}
\begin{table}
  \caption{LO and NLO total width in (GeV) and cross section with PDF errors
    in (fb) vs.\ $W^\prime_R$ mass for $pp\to W^{\prime}_{R}\to t\bar b\,
    (\bar t b)$ at $\sqrt{S}=14$ TeV, where the decay to leptons is not
    allowed.
    \label{tab:sigwprf}}
\begin{ruledtabular}
\begin{tabular}{rrrrrr@{\hspace*{-1.5em}}lr@{\hspace*{-1.5em}}l}
Mass (GeV) & $\Gamma_{\mathrm LO}$ (GeV)
& $\sigma^{t\bar b}_{\mathrm LO}\;\mathrm{(fb)}$
&$\sigma^{\bar t b}_{\mathrm LO}\;\mathrm{(fb)}$
& $\Gamma_{\mathrm NLO}$ (GeV)
&\multicolumn{2}{c}{$\sigma^{t\bar b}_{\mathrm NLO}\;\mathrm{(fb)}$}
&\multicolumn{2}{c}{$\sigma^{\bar t b}_{\mathrm NLO}\;\mathrm{(fb)}$}\\\hline
500 & 11.97 & 87300 & 48400 & 12.47 & 112000 & $^{+7600}_{-5200}$ & 62700 & $^{+5800}_{-1600}$\\
750 & 18.58 & 22700 & 11300 & 19.24 & 29000 & $^{+1400}_{-1800}$ & 14600 & $^{+700}_{-700}$\\
1000 & 25.07 & 7960 & 3580 & 25.88 & 9970 & $^{+960}_{-350}$ & 4680 & $^{+270}_{-300}$\\
1250 & 31.51 & 3350 & 1380 & 32.46 & 4210 & $^{+200}_{-420}$ & 1780 & $^{+170}_{-100}$\\
1500 & 37.93 & 1580 & 600 & 39.03 & 1950 & $^{+120}_{-180}$ & 779 & $^{+67}_{-64}$\\
1750 & 44.33 & 805 & 284 & 45.58 & 984 & $^{+58}_{-116}$ & 369 & $^{+33}_{-40}$\\
2000 & 50.72 & 432 & 142 & 52.11 & 514 & $^{+60}_{-42}$ & 186 & $^{+20}_{-17}$\\
2250 & 57.11 & 242 & 75.5 & 58.65 & 286 & $^{+27}_{-33}$ & 98.7 & $^{+11.0}_{-11.1}$\\
2500 & 63.49 & 139 & 41.4 & 65.17 & 163 & $^{+14.0}_{-26.3}$ & 54.3 & $^{+6.8}_{-6.8}$\\
2750 & 69.86 & 81.6 & 23.4 & 71.70 & 93.9 & $^{+13.0}_{-13.7}$ & 30.9 & $^{+4.2}_{-4.4}$\\
3000 & 76.24 & 48.8 & 13.6 & 78.22 & 55.5 & $^{+8.1}_{-9.1}$ & 18.1 & $^{+3.0}_{-2.8}$\\
3250 & 82.61 & 29.6 & 8.06 & 84.73 & 33.3 & $^{+5.3}_{-6.5}$ & 10.8 & $^{+1.8}_{-1.9}$\\
3500 & 88.99 & 18.1 & 4.88 & 91.23 & 20.2 & $^{+3.5}_{-4.3}$ & 6.57 & $^{+1.43}_{-1.02}$\\
3750 & 95.36 & 11.2 & 3.02 & 97.76 & 12.4 & $^{+2.5}_{-2.6}$ & 4.13 & $^{+0.84}_{-0.82}$\\
4000 & 101.7 & 7.05 & 1.90 & 104.3 & 7.72 & $^{+1.70}_{-1.82}$ & 2.62 & $^{+0.63}_{-0.49}$\\
4250 & 108.1 & 4.47 & 1.226 & 110.8 & 4.86 & $^{+1.15}_{-1.17}$ & 1.71 & $^{+0.41}_{-0.34}$\\
4500 & 114.5 & 2.87 & 0.806 & 117.3 & 3.13 & $^{+0.74}_{-0.80}$ & 1.13 & $^{+0.29}_{-0.21}$\\
4750 & 120.8 & 1.87 & 0.543 & 123.8 & 2.05 & $^{+0.52}_{-0.51}$ & 0.765 & $^{+0.205}_{-0.137}$\\
5000 & 127.2 & 1.24 & 0.375 & 130.3 & 1.37 & $^{+0.35}_{-0.33}$ & 0.535 & $^{+0.126}_{-0.104}$\\
5250 & 133.6 & 0.841 & 0.266 & 136.8 & 0.945 & $^{+0.239}_{-0.211}$ & 0.380 & $^{+0.091}_{-0.065}$\\
5500 & 139.9 & 0.586 & 0.194 & 143.3 & 0.671 & $^{+0.161}_{-0.136}$ & 0.278 & $^{+0.062}_{-0.044}$\\
5750 & 146.3 & 0.419 & 0.145 & 149.8 & 0.493 & $^{+0.103}_{-0.095}$ & 0.209 & $^{+0.042}_{-0.030}$\\
6000 & 152.7 & 0.309 & 0.112 & 156.3 & 0.371 & $^{+0.080}_{-0.054}$ & 0.161 & $^{+0.028}_{-0.022}$\\
6250 & 159.0 & 0.235 & 0.088 & 162.8 & 0.290 & $^{+0.050}_{-0.040}$ & 0.127 & $^{+0.020}_{-0.014}$\\
6500 & 165.4 & 0.184 & 0.071 & 169.3 & 0.231 & $^{+0.035}_{-0.027}$ & 0.102 & $^{+0.014}_{-0.010}$\\
6750 & 171.8 & 0.148 & 0.059 & 175.8 & 0.188 & $^{+0.026}_{-0.019}$ & 0.084 & $^{+0.009}_{-0.008}$\\
7000 & 178.1 & 0.121 & 0.049 & 182.3 & 0.157 & $^{+0.018}_{-0.014}$ & 0.070 & $^{+0.008}_{-0.006}$
\end{tabular}
\end{ruledtabular}
\end{table}
\endgroup

\begingroup
\renewcommand{\arraystretch}{2}
\begin{table}
  \caption{LO and NLO total width in (GeV) and cross section with PDF errors
    in (fb) vs.\ $W^\prime_L$ mass for $pp\to W^{\prime}_{L}\to t\bar b\,
    (\bar t b)$ at $\sqrt{S}=14$ TeV, where the decay to leptons is allowed,
    but no interference is included.
    \label{tab:sigwplf}}
\begin{ruledtabular}
\begin{tabular}{rrrrrr@{\hspace*{-1.5em}}lr@{\hspace*{-1.5em}}l}
Mass (GeV) & $\Gamma_{\mathrm LO}$ (GeV)
& $\sigma^{t\bar b}_{\mathrm LO}\;\mathrm{(fb)}$
&$\sigma^{\bar t b}_{\mathrm LO}\;\mathrm{(fb)}$
& $\Gamma_{\mathrm NLO}$ (GeV)
&\multicolumn{2}{c}{$\sigma^{t\bar b}_{\mathrm NLO}\;\mathrm{(fb)}$}
&\multicolumn{2}{c}{$\sigma^{\bar t b}_{\mathrm NLO}\;\mathrm{(fb)}$}\\\hline
500 & 16.21 & 64100 & 35600 & 16.71 & 84300 & $^{+2000}_{-6900}$ & 47200 & $^{+2000}_{-3200}$\\
750 & 24.95 & 16800 & 8370 & 25.60 & 21700 & $^{+1200}_{-1000}$ & 11000 & $^{+600}_{-500}$\\
1000 & 33.56 & 5950 & 2680 & 34.36 & 7610 & $^{+330}_{-520}$ & 3510 & $^{+270}_{-200}$\\
1250 & 42.12 & 2510 & 1040 & 43.07 & 3190 & $^{+120}_{-350}$ & 1350 & $^{+110}_{-90}$\\
1500 & 50.66 & 1190 & 452 & 51.76 & 1480 & $^{+90}_{-140}$ & 594 & $^{+42}_{-51}$\\
1750 & 59.18 & 606 & 215 & 60.43 & 742 & $^{+48}_{-69}$ & 282 & $^{+21}_{-28}$\\
2000 & 67.69 & 326 & 108 & 69.08 & 395 & $^{+28}_{-44}$ & 143 & $^{+13}_{-16}$\\
2250 & 76.20 & 183 & 57.5 & 77.74 & 219 & $^{+17}_{-31}$ & 76.1 & $^{+7.2}_{-9.2}$\\
2500 & 84.70 & 105 & 31.7 & 86.39 & 124 & $^{+16}_{-14}$ & 41.9 & $^{+5.3}_{-5.0}$\\
2750 & 93.20 & 62.2 & 18.1 & 95.03 & 72.7 & $^{+8.0}_{-12.0}$ & 24.0 & $^{+3.4}_{-3.2}$\\
3000 & 101.7 & 37.4 & 10.6 & 103.7 & 42.9 & $^{+6.0}_{-6.7}$ & 14.2 & $^{+2.1}_{-2.2}$\\
3250 & 110.2 & 22.8 & 6.33 & 112.3 & 25.9 & $^{+4.1}_{-4.6}$ & 8.52 & $^{+1.52}_{-1.25}$\\
3500 & 118.7 & 14.1 & 3.88 & 120.9 & 15.9 & $^{+2.6}_{-3.3}$ & 5.28 & $^{+0.98}_{-0.90}$\\
3750 & 127.2 & 8.82 & 2.43 & 129.6 & 9.86 & $^{+1.87}_{-2.07}$ & 3.32 & $^{+0.73}_{-0.54}$\\
4000 & 135.7 & 5.59 & 1.55 & 138.2 & 6.23 & $^{+1.24}_{-1.38}$ & 2.16 & $^{+0.44}_{-0.41}$\\
4250 & 144.2 & 3.59 & 1.02 & 146.8 & 3.98 & $^{+0.87}_{-0.90}$ & 1.42 & $^{+0.30}_{-0.27}$\\
4500 & 152.6 & 2.34 & 0.683 & 155.5 & 2.61 & $^{+0.58}_{-0.62}$ & 0.958 & $^{+0.225}_{-0.168}$\\
4750 & 161.1 & 1.55 & 0.469 & 164.1 & 1.74 & $^{+0.40}_{-0.39}$ & 0.662 & $^{+0.155}_{-0.110}$\\
5000 & 169.6 & 1.05 & 0.331 & 172.7 & 1.19 & $^{+0.28}_{-0.26}$ & 0.471 & $^{+0.103}_{-0.077}$\\
5250 & 178.1 & 0.731 & 0.240 & 181.4 & 0.842 & $^{+0.186}_{-0.167}$ & 0.343 & $^{+0.068}_{-0.056}$\\
5500 & 186.6 & 0.521 & 0.179 & 190.0 & 0.612 & $^{+0.126}_{-0.113}$ & 0.256 & $^{+0.048}_{-0.036}$\\
5750 & 195.1 & 0.382 & 0.136 & 198.6 & 0.459 & $^{+0.082}_{-0.077}$ & 0.195 & $^{+0.035}_{-0.024}$\\
6000 & 203.6 & 0.288 & 0.107 & 207.2 & 0.353 & $^{+0.056}_{-0.054}$ & 0.153 & $^{+0.023}_{-0.018}$\\
6250 & 212.1 & 0.223 & 0.085 & 215.9 & 0.277 & $^{+0.043}_{-0.031}$ & 0.122 & $^{+0.017}_{-0.012}$\\
6500 & 220.6 & 0.177 & 0.070 & 224.5 & 0.225 & $^{+0.029}_{-0.024}$ & 0.100 & $^{+0.011}_{-0.010}$\\
6750 & 229.0 & 0.144 & 0.058 & 233.1 & 0.185 & $^{+0.020}_{-0.019}$ & 0.082 & $^{+0.008}_{-0.007}$\\
7000 & 237.5 & 0.119 & 0.048 & 241.7 & 0.155 & $^{+0.014}_{-0.015}$ & 0.069 & $^{+0.007}_{-0.005}$
\end{tabular}
\end{ruledtabular}
\end{table}
\endgroup

\begingroup
\renewcommand{\arraystretch}{2}
\begin{table}
  \caption{LO and NLO total width in (GeV) and cross section 
    in (fb) vs.\ $W^\prime_R$ mass for $pp\to W^{\prime}_{R}\to t\bar b\,
    (\bar t b)$ at $\sqrt{S}=33$ TeV, where the decay to leptons is not
    allowed.
    \label{tab:sigwprt}}
\begin{ruledtabular}
\begin{tabular}{rrrrrrr}
Mass (GeV) & $\Gamma_{\mathrm LO}$ (GeV)
& $\sigma^{t\bar b}_{\mathrm LO}\;\mathrm{(fb)}$
&$\sigma^{\bar t b}_{\mathrm LO}\;\mathrm{(fb)}$
& $\Gamma_{\mathrm NLO}$ (GeV)
&$\sigma^{t\bar b}_{\mathrm NLO}\;\mathrm{(fb)}$
&$\sigma^{\bar t b}_{\mathrm NLO}\;\mathrm{(fb)}$\\\hline
500 & 11.97 & 281000 & 183000 & 12.47 & 345000 & 229000\\
1000 & 25.07 & 31100 & 17900 & 25.88 & 38400 & 22900\\
1500 & 37.93 & 7610 & 3960 & 39.03 & 9560 & 5030\\
2000 & 50.72 & 2660 & 1270 & 52.11 & 3290 & 1620\\
2500 & 63.49 & 1130 & 496 & 65.17 & 1410 & 638\\
3000 & 76.24 & 543 & 221 & 78.22 & 665 & 281\\
4000 & 101.7 & 157 & 56.3 & 104.3 & 188 & 72.1\\
5000 & 127.2 & 54.3 & 17.5 & 130.3 & 63.7 & 22.6\\
6000 & 152.7 & 20.8 & 6.23 & 156.3 & 24.0 & 8.09\\
7000 & 178.1 & 8.51 & 2.43 & 182.3 & 9.53 & 3.18\\
8000 & 203.6 & 3.64 & 1.02 & 208.4 & 4.00 & 1.35\\
9000 & 229.1 & 1.62 & 0.457 & 234.4 & 1.76 & 0.609\\
10000 & 254.5 & 0.750 & 0.220 & 260.4 & 0.812 & 0.297\\
11000 & 280.0 & 0.363 & 0.114 & 286.3 & 0.395 & 0.155\\
12000 & 305.4 & 0.187 & 0.064 & 312.3 & 0.208 & 0.087\\
13000 & 330.9 & 0.104 & 0.039 & 338.3 & 0.119 & 0.053\\
14000 & 356.4 & 0.063 & 0.025 & 364.3 & 0.074 & 0.035\\
15000 & 381.8 & 0.041 & 0.018 & 390.3 & 0.051 & 0.024
\end{tabular}
\end{ruledtabular}
\end{table}
\endgroup

\begingroup
\renewcommand{\arraystretch}{2}
\begin{table}
  \caption{LO and NLO total width in (GeV) and cross section with PDF errors
    in (fb) vs.\ $W^\prime_L$ mass for $pp\to W^{\prime}_{L}\to t\bar b\,
    (\bar t b)$ at $\sqrt{S}=33$ TeV, where the decay to leptons is allowed,
    but no interference is included.
    \label{tab:sigwplt}}
\begin{ruledtabular}
\begin{tabular}{rrrrrrr}
Mass (GeV) & $\Gamma_{\mathrm LO}$ (GeV)
& $\sigma^{t\bar b}_{\mathrm LO}\;\mathrm{(fb)}$
&$\sigma^{\bar t b}_{\mathrm LO}\;\mathrm{(fb)}$
& $\Gamma_{\mathrm NLO}$ (GeV)
&$\sigma^{t\bar b}_{\mathrm NLO}\;\mathrm{(fb)}$
&$\sigma^{\bar t b}_{\mathrm NLO}\;\mathrm{(fb)}$\\\hline
500 & 16.21 & 206000 & 134000 & 16.71 & 254000 & 172000\\
1000 & 33.56 & 23200 & 13400 & 34.36 & 29600 & 17100\\
1500 & 50.66 & 5700 & 2970 & 51.76 & 7210 & 3850\\
2000 & 67.69 & 1990 & 952 & 69.08 & 2490 & 1220\\
2500 & 84.70 & 848 & 374 & 86.39 & 1050 & 488\\
3000 & 101.7 & 408 & 167 & 103.7 & 504 & 216\\
4000 & 135.7 & 119 & 42.8 & 138.2 & 143 & 55.7\\
5000 & 169.6 & 41.2 & 13.5 & 172.7 & 48.5 & 17.4\\
6000 & 203.6 & 15.9 & 4.83 & 207.2 & 18.4 & 6.32\\
7000 & 237.5 & 6.56 & 1.91 & 241.7 & 7.43 & 2.51\\
8000 & 271.5 & 2.85 & 0.818 & 276.2 & 3.18 & 1.09\\
9000 & 305.4 & 1.29 & 0.377 & 310.7 & 1.43 & 0.506\\
10000 & 339.4 & 0.612 & 0.187 & 345.2 & 0.676 & 0.253\\
11000 & 373.3 & 0.306 & 0.100 & 379.7 & 0.340 & 0.137\\
12000 & 407.3 & 0.164 & 0.058 & 414.1 & 0.186 & 0.079\\
13000 & 441.2 & 0.095 & 0.036 & 448.6 & 0.111 & 0.050\\
14000 & 475.1 & 0.059 & 0.024 & 483.1 & 0.071 & 0.033\\
15000 & 509.1 & 0.040 & 0.017 & 517.5 & 0.049 & 0.023
\end{tabular}
\end{ruledtabular}
\end{table}
\endgroup

Many models that attempt to explain electroweak symmetry breaking
predict the existence of $W^\prime$ bosons.  This white paper provides
the LO and NLO cross sections and total widths necessary to determine
the model-independent reach at 14~TeV and 33~TeV $pp$ colliders.
Since resonant production is expected up to $\sim 1/3$ of the collider
energy, a 33~TeV $pp$ collider should be able to observe or exclude
most perturbative models with $W^\prime$ bosons with masses below
$10$~TeV.

\begin{acknowledgments}
  This work is supported by the U.S.\ Department of Energy under
  Contract No.\ DE-SC0008347.
\end{acknowledgments}

\end{document}